\documentclass[aps,12pt,superscriptaddress,amsfonts,amssymb,amsmath]{revtex4}
\pdfoutput=1

\usepackage{graphicx}
\usepackage{epsfig}
\usepackage{epstopdf}
\usepackage{natbib}
\usepackage{xcolor}

\usepackage{verbatim}
\usepackage{empheq}
\usepackage{dsfont}
\usepackage{caption}
\usepackage{subcaption}
\usepackage{tabulary}

\begin{document}

\title{Generation of scale-free assortative networks via Newman rewiring for simulation of diffusion phenomena}

\author{L.\ Di Lucchio \footnote{Email address: Laura.DiLucchio@unibz.it}}
\affiliation{Free University of Bozen-Bolzano \\ Faculty of Engineering \\ I-39100 Bolzano, Italy}

\author{G.\ Modanese \footnote{Email address: Giovanni.Modanese@unibz.it}}
\affiliation{Free University of Bozen-Bolzano \\ Faculty of Engineering \\ I-39100 Bolzano, Italy}
\date{\today}

\linespread{0.9}

\begin{abstract}
By collecting and expanding several numerical recipes developed in previous work, we implement an object-oriented Python code, based on the \texttt{networkX} library, for the realization of the configuration model and Newman rewiring. The software can be applied to any kind of network and ``target'' correlations, but it is tested with focus on scale-free networks and assortative correlations. For generating the degree sequence we use the method of ``random hubs'', which gives networks with minimal fluctuations. For the assortative rewiring we use the simple Vazquez-Weigt matrix as a test in the case of random networks; since it appears to be not effective in the case of scale-free networks we then turn to another recipe which generates matrices with decreasing off-diagonal elements. The rewiring procedure is important also at the theoretical level, in order to test which types of statistically acceptable correlations can actually be realized in concrete networks. From the applicative point of view, its main use is in the construction of correlated networks for the solution of dynamical or diffusion processes by analyzing the evolution of single nodes, i.e., beyond the Heterogeneous Mean Field approximation. As an example, we report on an application to the Bass diffusion model, with calculations of the time $t_{max}$ of the diffusion peak. The same networks can additionally be exported in environments for agent-based simulations like NetLogo.

\end{abstract}

\maketitle

\section{Introduction}

The configuration model is a well-known method for constructing uncorrelated networks having an assigned degree distribution. Such networks can be used, for example, as a connectivity environment for the simulation of dynamical processes or diffusion phenomena. In particular, one is often interested into scale-free networks, i.e., with degree distribution of the form $P(k)=c_\gamma k^{-\gamma}$, where $P(k)$ is the probability that a randomly chosen node has degree $k$. There are standard commands in software packages like Python's \texttt{networkX} for generating samples from a scale-free distribution, and for performing on the corresponding ``stubs'' of nodes a wiring procedure uncorrelated from the degree. In Sect.\ \ref{test-CL} of this work we briefly recall this method and we compare its accuracy end efficiency with another method originally proposed in \cite{bertotti2019configuration} (``random hubs'' method) and here implemented using the object-based functions of \texttt{networkX}. (All codes are available at \texttt{https://github.com/Ladilu/python-bass-accessible}.)

Much less is known about the possibility of generating scale-free networks with assigned degree correlations, especially assortative networks, which are usually employed to represent social networks, financial and economic networks, and also some traffic networks and technological networks \cite{noldus2015assortativity}. For this purpose it is possible in principle to apply to an uncorrelated scale-free network the Newman rewiring procedure with assortative target correlations \cite{newman2003mixing}. It is not easy to tell in advance, however, if and when this procedure will be successful. The main purpose of this paper is to explore this issue and find some useful common principles and numerical recipes.

At which level can degree correlations be assigned in a network? Given a statistical degree correlation matrix $P(h|k)$ having the correct properties of positivity, normalization and pseudo-symmetry (and there are many ways of constructing such matrices), when can we say that scale-free networks with those correlations exist, or at least that a statistical ensemble of networks exists, which exhibit those correlations at an average level?

Answering such questions is important for the sake of network theory itself, but also because in the so-called Heterogeneous Mean Field (HMF) approximation \cite{boguna2003epidemic,vespignani2012modelling,pastor2015epidemic} several general results concerning diffusion processes have been obtained postulating certain statistical correlations, or just requiring that the function $\bar{k}_{nn}(k)$ derived from the full $P(h|k)$ has certain properties. We recall that $\bar{k}_{nn}(k)$ represents the average degree of the neighbours of a node of degree $k$:
\begin{equation}
    \bar{k}_{nn}(k)=\sum_{h=1}^n hP(h|k)
    \label{def-knn}
\end{equation}
In general, networks with different correlations $P(h|k)$ may have the same $\bar{k}_{nn}(k)$, see \cite{bertotti2021diagonal}.
Rigorous results about the convergence of the $\bar{k}_{nn}(k)$ function in random graphs with given joint degree distribution of neighbor nodes and in the configuration model have been given by \cite{yao2018average}.

A typical example is the assortative matrix by Vazquez-Weigt \cite{vazquez2003computational}. It has the form
\begin{equation}
    P^{Vaz}(h|k)=(1-r)\frac{hP(h)}{\langle k \rangle} +r\delta_{hk}
\end{equation}
and its $\bar{k}_{nn}(k)$ function is
\begin{equation}
    \bar{k}^{Vaz}_{nn}(k)=(1-r)\frac{\langle k^2 \rangle}{\langle k \rangle} + rk
    \label{knn-vaz}
\end{equation}
The Newman assortativity coefficient $r$ enters explicitly in the definition (in general $r$ spans the range $[-1,1]$, but here $0 \leq r \leq 1$). Notice that this $\bar{k}_{nn}(k)$ is linear but does not start from the origin. Using the correlation matrix above one can quickly write diffusion equations in HMF approximation, namely a system of $n$ coupled differential equations, one for each degree class, which e.g.\ for the Bass diffusion model have the form
\begin{equation}
    \frac{dG_i(t)}{dt} = [1-G_i(t)] \left[  p+iq \sum_{h=1}^n P(h|i)G_h(t)  \right], \qquad i=1,...,n
\end{equation}

With $p=0$ (no ``publicity'' or ``broadcast'' term) these equations reduce to those of the SI (``Susceptible-Infected'') epidemic model. However, it turns out that it is not always possible to build real scale-free networks having this kind of correlations.

The outline of the work is the following. In Sect.\ \ref{newman} we introduce Newman's assortative rewiring by applying it to the case of Erd\"os-Renyi random networks. In this case the technique works quite well already with the  assortative matrix by Vazquez-Weigt. In Sect.\ \ref{ba} we show that the rewiring technique fails for Barabasi-Albert networks, partly due to their intrinsic correlations, but also to their scale-free degree distribution, as confirmed in the following, when other scale-free networks are considered. In Sect.\ \ref{conf} we look at the general scale-free case with exponents between 2 and 3. For this we first need to implement in an efficient way the configuration model in order to build the uncorrelated networks from which the rewiring process begins. We note that the Chung-Lu method has some limitations under this respect, at least when the size of the networks is not very large. For this reason we introduce the method of the random hubs (Sect.\ \ref{hubs}). In Sect.\ \ref{results-sf} we apply the rewiring procedure in the general scale-free case and we find that it works quite well with some target assortative matrices which were defined in previous work; they do not have an explicit expression like the Vazquez-Weigt, but the construction algorithm is included in the present code. In Sect.\ \ref{num} we implement the network Bass model in first closure on the single nodes, i.e., with $N$ differential equations coupled via the adjacency matrix ($N$ is the number of nodes). We briefly discuss the outcome concerning the total diffusion curves and the peak time of the total diffusion rate, also comparing the uncorrelated and assortative case. In Sect.\ \ref{Netlogo-sim} we give a brief outline on future work in which we use \texttt{NetLogo} for agent-based simulations of the network Bass model (including extensions with modified network links and dynamics) implemented on the assortative networks built in this work.

\section{Newman rewiring for random networks}
\label{newman}

\begin{figure}[ht] 
\centering \includegraphics[width=0.6\columnwidth]{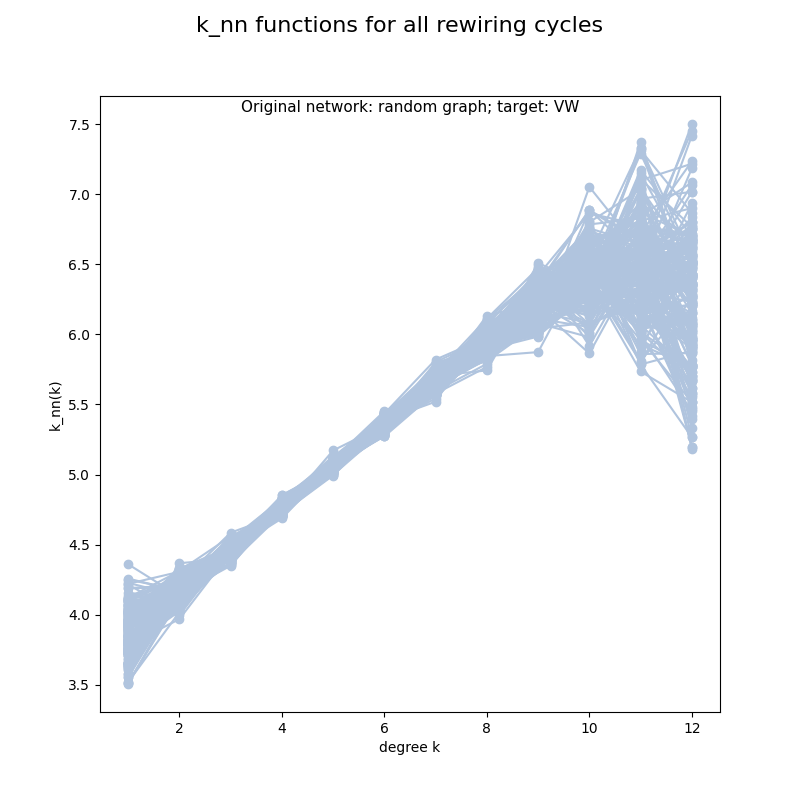}
\caption{
The $\bar{k}_{nn}$ functions of an ensemble of 200 networks obtained by Newman rewiring with target correlations of the Vazquez-Weigt type ($r=0.3$), starting from a random network of 4000 nodes, probability connection 0.001. The Newman coefficient of the ensemble is $r=0.20 \pm 0.01$. The giant component after the last rewiring is 97.9 \%. In each of the 200 rewiring sub-cycles, the number of accepted rewiring steps was about 218'000.
}
\label{rnd-cloud}
\end{figure}

\begin{figure}[ht] 
\centering \includegraphics[width=0.6\columnwidth]{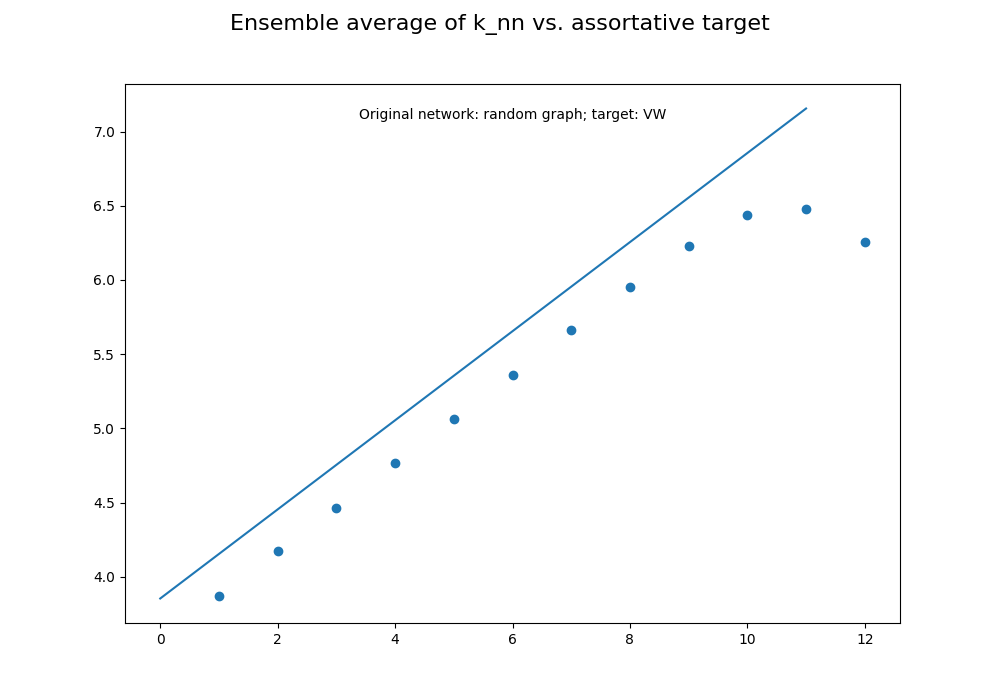}
\caption{
Average of the $\bar{k}_{nn}$ functions of the ensemble of random networks shown in Fig.\ \ref{rnd-cloud} (dotted plot). The continuous line shows the target $\bar{k}_{nn}(k)$ function (eq.\ (\ref{knn-vaz}), with $r=0.3$).
}
\label{rnd-mean}
\end{figure}

Before focusing on the case of scale-free networks, we notice that the rewiring procedure does work quite well for random networks. As seen from the examples in Figs.\ \ref{rnd-cloud}, \ref{rnd-mean} one obtains, as an average in the ensemble of rewired networks, a $\bar{k}_{nn}(k)$ function which is linear like the target, except for large $k$. This deviation at large $k$ was actually expected, as a consequence of the known ``structural disassortativity'' effect, due to the small number of hubs, which makes impossible for hubs to connect mainly to other hubs as required in principle by assortative correlations.

Let us briefly explain how the plots in Figs.\ \ref{rnd-cloud}, \ref{rnd-mean} are obtained. This allows us to first illustrate the rewiring method in a simplified case where the degree distribution is obtained with simple \texttt{networkX} commands, without using the configuration model as described later for the scale-free case. First we generate an uncorrelated random network with \texttt{G = nx.gnp\_random\_graph(N, p, seed=...)} assigning a certain number of nodes $N$ and a certain probability of connection $p$. Then we extract its degree sequence with \texttt{deg\_sequence=sorted((d for n, d in G.degree()), reverse=True)}. The degree sequence is passed to a custom-built function called \texttt{random\_reference} which performs a number of rewiring cycles fixed by the variable \texttt{num\_cycles} (typically a few hundreds for better statistics, although already after the first few cycles correlations become close to the target as far as possible). The number of rewiring steps in each cycle is defined by the function parameter \texttt{niter}, which should be adjusted in such a way that each link of the network is cut and re-connected at least 10 times in a cycle.

The general working principle of the Newman rewiring method is the following.
First one chooses at random in the list of the
links of the network two links $(a,b)$ and $(c,d)$, between nodes $a,b,c,d$, with excess degrees $A,B,C,D$. Then one computes the probability $E_1$ of these links according
to a ``target'' correlation matrix $e^0$, namely $E_1=e^0_{AB}e^0_{CD}$
and the analogous probability $E_2$ for the exchanged links, i.e.\ for the
couple $(a,c)$ and $(b,d)$.
After this, a sort of Metropolis-Monte Carlo criterion is applied: if $E_2>E_1$, then the rewiring is
performed with probability 1, otherwise it is performed with a probability proportional to the ratio $E_1/E_2$.

In our code the target correlations are defined in terms of the nodes degrees by assigning $P(h|k)$ before the rewiring cycles, and translated into the correlations $e^0_{ij}$ of the excess degrees with the formula $e^0_{ij}=P(i+1|j+1)(j+1)P(j+1)/\langle k \rangle$, written as \texttt{e0[i,j]=(Phk[i+1,j+1]*(j+1)*probability\_seq[j+1])/aver\_degree}. The array \texttt{probability\_seq} gives the normalized degree distribution of the random graph \texttt{G} and is obtained via the function \texttt{degree\_dist(G)}, which in turn uses the command \texttt{nx.degree\_histogram(G)}. Note that the indices $i$ and $j$ are in the range $0,...,n-1$, while the indices $h$ and $k$ are in the range $1,...,n$.

The $\bar{k}_{nn}(k)$ function of the target correlation matrix is also computed, according to the definition \ref{def-knn}, for comparison with the average $\bar{k}_{nn}(k)$ of the rewired ensemble. A counter variable returns for each sub-cycle the number of rewiring steps which have been accepted. (See typical values in Figs.\ \ref{rnd-cloud}, \ref{rnd-mean})

The choice of the edges for the rewiring is not performed by creating the edge list, but instead the module \texttt{discrete\textunderscore sequence} is called. The latter returns a sample sequence of length $n$ from the discrete cumulative distribution of the degrees. Then two numbers are picked from the sequence; if they are the same, the code skips. The choice among the neighbours for the rewiring is made  by means of the built-in python module \texttt{random}, which allows to generate and choose random numbers in a list. The vertices are selected so that they are different from each other. Before finally performing the rewiring, after the condition is proven, another control is made, in order to check whether the edges that are going to be added already exist; in fact, they must not be already present in the graph, or the \texttt{networkX} package will not perform the rearrangement as prescribed.

At the end of each rewiring sub-cycle the $r$ coefficient of the network is computed and stored for a final evaluation of the average and standard deviation of $r$ over all sub-cycles. As we shall see, this gives only a rough ``integral'' check of the convergence of the degree correlations to the target correlations. 

Moreover, at the end of each sub-cycle the $\bar{k}_{nn}(k)$ function of the current rewired network \texttt{G2} is computed with the command \texttt{knn=nx.average\_degree\_connectivity(G2,...)} and accumulated into a dictionary \texttt{knn1= dict(sorted(knn.items()))} for subsequent evaluation of its final average. All the graphs of $\bar{k}_{nn}(k)$ for each sub-cycle are plotted together in a ``cloud'' graph which gives a visual impression of their fluctuations. Finally, the ensemble average of $\bar{k}_{nn}(k)$ and the target $\bar{k}_{nn}(k)$ are plotted for comparison. The final size of the giant connected component is also computed.

\section{Newman rewiring for Barabasi-Albert networks}
\label{ba}

\begin{figure}[ht] 
\centering \includegraphics[width=0.6\columnwidth]{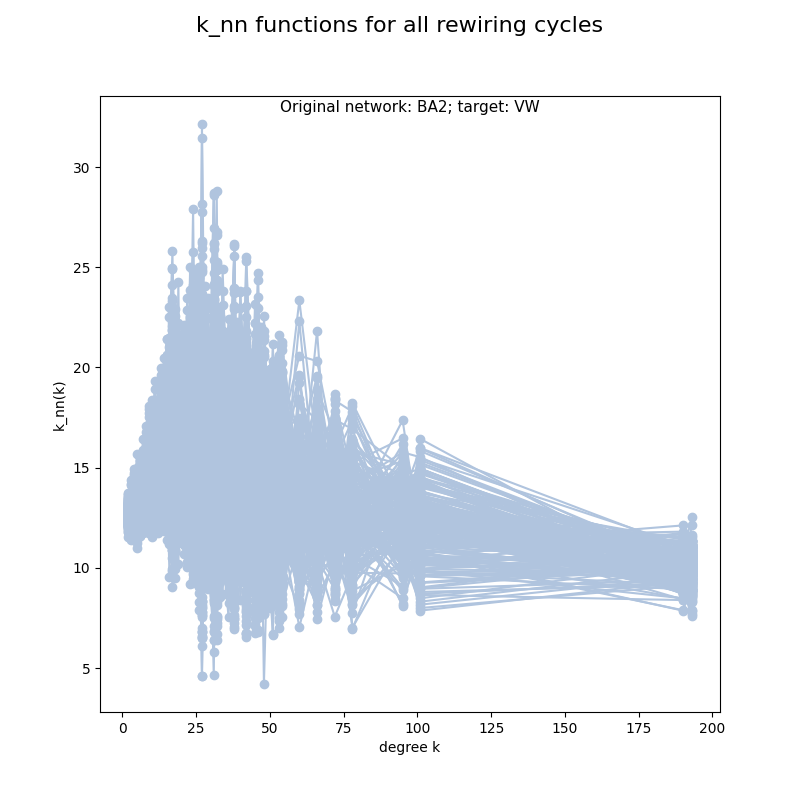}
\caption{
The $\bar{k}_{nn}$ functions of an ensemble of 200 networks obtained by Newman rewiring with target correlations of the Vazquez-Weigt type ($r=0.3$), starting from a BA-2 network of 5000 nodes. The Newman coefficient of the ensemble is $r=-0.07 \pm 0.06$. The giant component after the last rewiring is 100 \%. In each of the 200 rewiring sub-cycles, the number of accepted rewiring steps was about 192'000.
}
\label{ba-cloud}
\end{figure}

\begin{figure}[ht] 
\centering \includegraphics[width=0.6\columnwidth]{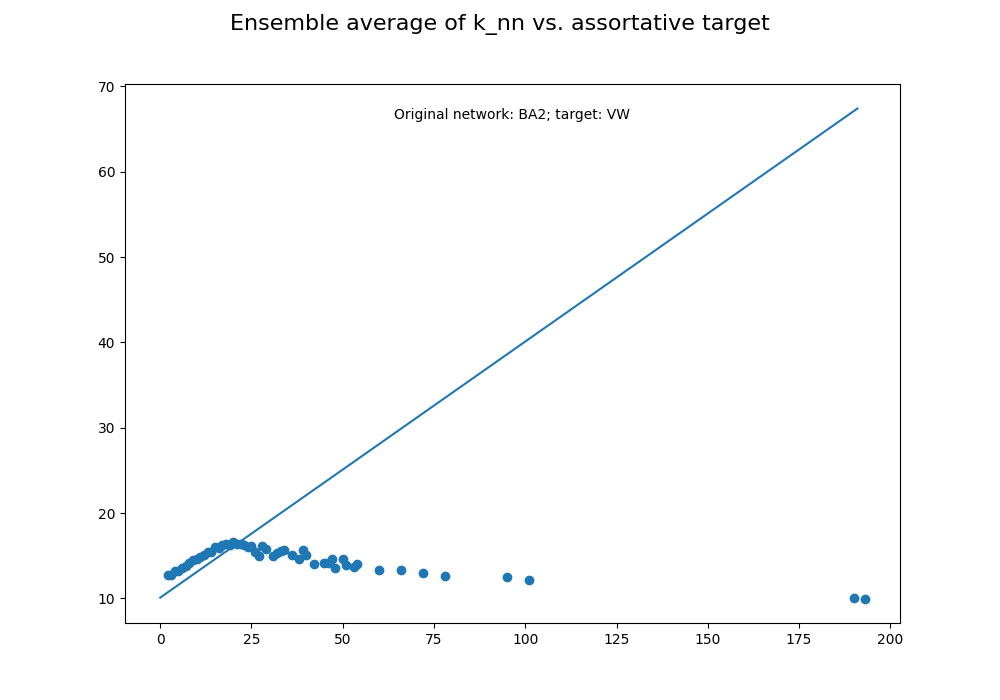}
\caption{
Average of the $\bar{k}_{nn}$ functions of the ensemble of BA networks shown in Fig.\ \ref{ba-cloud} (dotted plot). The continuous line shows the target $\bar{k}_{nn}(k)$ function (eq.\ (\ref{knn-vaz}), with $r=0.3$).
}
\label{ba-mean}
\end{figure}

A rewiring procedure similar to the one starting from random networks can be applied to Barabasi-Albert (BA) networks generated by preferential attachment to $\alpha$ existing nodes with the method \texttt{G=nx.barabasi\_albert\_graph(N,alpha)}. In this case, however, the rewired network turns out to be not assortative, and so the procedure fails. See Figs.\ \ref{ba-cloud}, \ref{ba-mean}. This is due to a limitation of the Newman rewiring when applied to scale-free networks (see Sect.\ \ref{results-sf}) and, in addition, to the fact that BA networks have intrinsic degree correlations which cannot apparently adapt well to the Vazquez-Weigt correlation matrix. In fact, although the $r$ coefficient of BA networks is close to zero, their $\bar{k}_{nn}(k)$ function is decreasing at small degrees and increasing at large degrees. 

We notice in this regard that the Newman rewiring works quite well, on the contrary, when one starts from an uncorrelated scale-free network with $\gamma=3$ and takes the BA correlation matrix as target \cite{bertotti2019configuration}.

The matching between target correlations and correlations of the rewired network is usually assessed by comparing the respective $\bar{k}_{nn}$ functions, at least for degrees which are not too large (the structural disassortativity prevents any real $\bar{k}_{nn}$ to increase indefinitely like a theoretical target assortative $\bar{k}_{nn}$, because there are simply not enough hubs in a scale-free network for connecting hubs preferably with hubs). A common phenomenon occurring in scale-free networks, however, is the strong fragmentation of the rewired network, especially if the smallest degree is $k_{min}=1$. In that case, the nodes with $k=1$ represent a vast majority of all nodes, and if $P(1|1)\neq 0$ in the target correlations, then a large number of isolated couples are formed; if $P(2|1)\neq 0$ a large number of isolated triples, and so on, leaving a small principal connected component. This phenomenon can be observed using the method \texttt{G2.subgraph(max(nx.connected\_components(G2), key=len))}. The situation improves markedly if the minimum degree is equal to 2. In this case, the principal component is really a ``giant'' component, close to 100\% of all nodes.

\section{Configuration model for scale-free networks}
\label{conf}

We are now going to illustrate a method for constructing uncorrelated scale-free networks with given exponent $2<\gamma\le 3$ and for applying Newman rewiring to them in order to approach some target assortative correlations as well as possible. It is known that many relevant real networks are assortative, in particular social and economic networks; it is therefore very useful, in order to simulate dynamical processes on this kind of networks, to generate them through some algorithm. This issue is also interesting in itself, for ``foundational'' purposes, in order to understand if the scale-free property is fully compatible with assortative (or possibly disassortative) correlations.

The established paradigm of scale-free complex networks has been recently re-discussed \cite{broido2019scale,holme2019rare} and it turns out that in many cases it is impossible, due to statistical errors and fit uncertainties, to state whether certain networks are scale-free or not. Since the preferential attachment method leads to an exponent $\gamma=3$, for the generation of scale-free networks with a different exponent it is usually necessary to employ the configuration model. In Sect.\ \ref{bm2-matrices} we shall present a novel alternative method \cite{bertotti2019evaluation} in which the excess degree correlations $e_{jk}$ are assigned at the beginning and a good approximation of scale-free degree distributions is obtained as a consequence.

\subsection{Test of the implementation of the Chung and Lu model}
\label{test-CL}

Before describing our implementation of the configuration model, we will discuss the limitations of a method already available in \texttt{networkX} and based on the model by Chung and Lu \cite{chung2002connected,miller2011efficient}. This is a weaker version of the configuration model, treated also by Newman in his book \cite{newman2010networks} in comparison to the standard configuration model with ``fixed degree distribution'', in which it is possible to prove analytically several general properties.

Suppose we want to build a network with $N$ nodes, starting from a list of $N$ values for the degrees of the nodes, randomly extracted from a power law distribution with given exponent $\gamma$. It is possible to generate such a list with various stochastic methods, in a similar way as for the generation of samples of the normal distribution or of other known distributions. A general procedure makes use of a probability
transformation method. One defines first a vector $F_k =\sum_{j=1}^kP(j)$
where $k = 1, ... , n$ and $P(j)$ denotes the normalized degree distribution (the power law in our case, with $n$ the maximum degree). The values of
$F_k$ define breakpoints of the unit interval $(0,1)$. After generating a random number $\xi$ in
this interval, a new node is introduced into the list with degree $k$ if $F_{k-1} <\xi< F_k$, and the procedure is repeated $N$ times. 

In \texttt{networkX} there is an auxiliary function which does just this, namely \texttt{S=nx.utils.powerlaw\_sequence(N, gamma)}. The list of values obtained is real, in the interval $(1,+\infty)$. For transforming this into a list of degrees, a proper rounding to integers is necessary. It can be checked that for large $N$ the degree succession obtained respects well the scale-free criterion of the ratio of probabilities, namely $P(k_1)/P(k_2)=(k_2/k_1)^\gamma$. 

In the Chung and Lu method (\texttt{networkX} command: \texttt{G = nx.expected\_degree\_graph(S, selfloops=False)}) one actually starts from the not-rounded scale-free degree succession and connects nodes $i$ and $j$ having degrees $k_i$ and $k_j$ ($i,j=1,...,N$) with probability proportional to $k_ik_j$. It is possible to prove that the network obtained must be scale-free in the limit $N\to \infty$. The algorithm by Miller and Hagberg employed by \texttt{networkX} \cite{miller2011efficient} runs in an expected time of order $N$. It is clear that the exact degrees of the nodes cannot be assigned a priori like in the standard configuration model, but become random variables themselves.

Our numerical tests of this method (code at \texttt{https://github.com/Ladilu/python-bass-accessible}) give, at least for values $N$ up to $10^5$, distributions of the final degrees which deviate significantly from the scale-free criterion (while the initial degree list, for the same values of $N$, is accurately scale-free). 

\subsection{Random hubs method}
\label{hubs}

Therefore we chose to implement the standard configuration model in a custom function named \texttt{configuration\_model}, in which a multigraph is constructed starting from the assigned degree sequence. In this work we only made use of the code for undirected network. Once the degree sequence is given, a list containing all nodes is built, each node being repeated as many times as its degree determines. Then the list is reshuffled by means of the \texttt{random} module functions and split in half. The vertices for the edges are picked in an ordered way from the two halves of the list, so that the method cannot go back and accidentally select one node more times than allowed. This is possible with the built-in methods of the Python \texttt{itertools} package, which are programmed for operating on iterable objects such as arrays and lists. The edges are finally added to the graph until the list is finished, in such a way that the original degrees of the nodes are preserved. 

This function works with any assigned degree sequence. For example, it is also possible to get from \texttt{networkX} a random graph or a BA network like in the rewiring examples previously discussed, then extract their degree sequence and apply the configuration model function to it, obtaining an uncorrelated network with the same degree sequence (of course, the random graph should already be  uncorrelated). In order to build a scale-free network one could apply the function \texttt{configuration\_model} to a degree sequence generated with the command \texttt{S=nx.utils.powerlaw\_sequence(N, gamma)} as described above, but we chose instead to define a scale-free degree sequence with ``minimal fluctuations'' through a technique that we call the ``random hubs method''. This allows to build networks of relatively small size which are as close as possible to an ideal scale-free distribution, and works as follows.

If the average number of nodes
with degree $k$ is smaller than 1, i.e., $NP(k) = X < 1$, then a node with this degree
will be created with probability $X$. Extending the procedure to all
degrees, a random variable $\xi \in (0, 1)$ is generated for each value of $k$, and then denoting by $Int(NP(k))$ the integer part of $NP(k)$ and by $Dec(NP(k))$  its decimal part, one sets the number $N_k$ of nodes with degree $k$ to
$N_k = Int(NP(k))$ if $\xi > Dec(NP(k))$ and $N_k = Int(NP(k)) + 1$ if $\xi < Dec(NP(k))$. The total
number of nodes is therefore not fixed, with random variations of 1 for each degree.

\begin{figure}[ht] 
\centering \includegraphics[width=0.6\columnwidth]{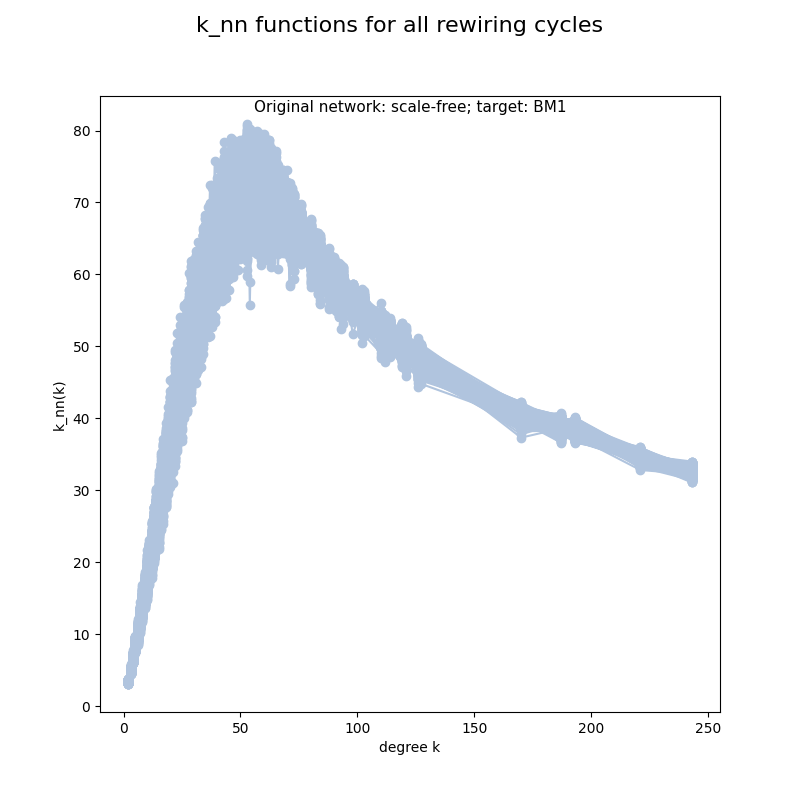}
\caption{
The $\bar{k}_{nn}$ functions of an ensemble of 400 networks obtained by Newman rewiring with target correlations of the BM1 type, starting from an uncorrelated scale-free network of 10000 nodes, $\gamma=2.5$. The Newman coefficient of the ensemble is $r=0.337 \pm 0.002$. The giant component after the last rewiring is 100 \%. In each of the 400 rewiring sub-cycles, the number of accepted rewiring steps was about 145'000.
}
\label{bm-cloud}
\end{figure}

\begin{figure}[ht] 
\centering \includegraphics[width=0.6\columnwidth]{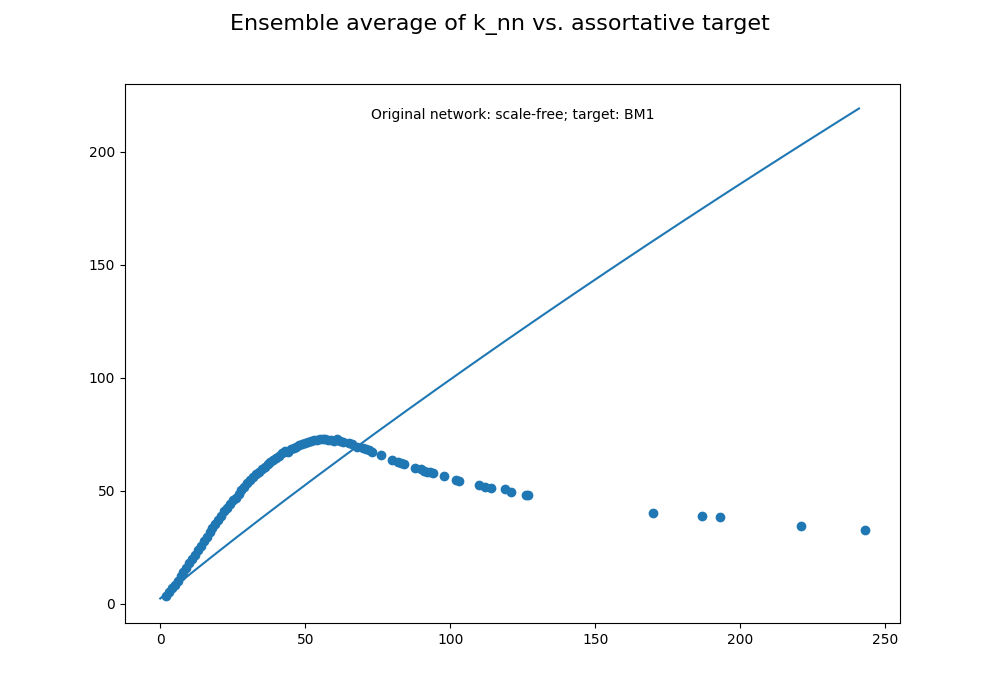}
\caption{
Average of the $\bar{k}_{nn}$ functions of the ensemble of random networks shown in Fig.\ \ref{bm-cloud} (dotted plot). The continuous line shows the target $\bar{k}_{nn}(k)$ function.
}
\label{bm-mean}
\end{figure}

\begin{figure}[ht] 
\centering \includegraphics[width=0.6\columnwidth]{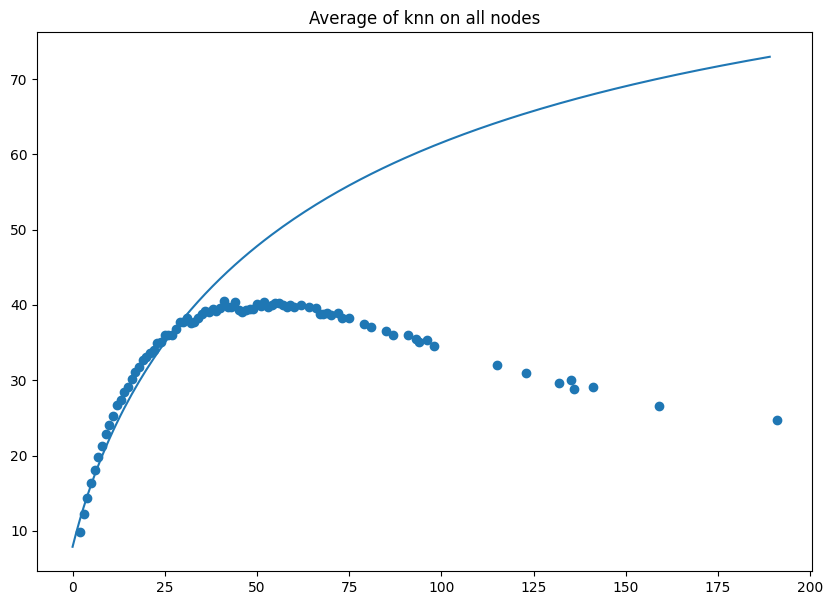}
\caption{
Average of the $\bar{k}_{nn}$ functions of an ensemble of 200 networks obtained by Newman rewiring with target correlations of the type $e_{jk}=c_\delta/(1+j+k)^\delta$, with $\delta=2.2$, starting from an almost scale-free uncorrelated network ($\gamma \simeq 2.42$) with degree distribution derived from $e_{jk}$. The number of nodes is $N=5000$. The Newman coefficient of the ensemble is $r=0.176 \pm 0.005$. The giant component after the last rewiring is 99.9 \%. In each of the 200 rewiring sub-cycles, the number of accepted rewiring steps was about 280'000.
}
\label{ejk-mean}
\end{figure}

With the random hubs method we thus obtain a list of values (called \texttt{s1} in the code) which shows how many nodes of each degree, starting from $k_{min}$, must be present in the network in order to satisfy the power law -- except for the minimal fluctuations mentioned above. For the highest degrees, most values will be zero, because only a few hubs will actually be present. This is what one also observes in the degree sequence of real scale-free networks, e.g.\ BA networks: some hubs, with random degrees, are present, and the remaining high degrees are missing.

The maximum degree $n$ considered for the list \texttt{s1} is connected to the number of nodes $N$ through the Dorogotsev-Mendes relation. After the list has been built, it is necessary to check that the total number is even, otherwise the random wiring procedure cannot connect all nodes respecting their degrees. From the list \texttt{s1} another list is built with the command \texttt{list\_of\_nodes = \_to\_stublist\_kmin(s1, kmin)}, giving the degree sequence of the nodes. For example, if we are generating a network with exponent $\gamma=2$ (this extreme value of $\gamma$ is normally excluded, and taken here only for simplicity) and there are 100 nodes with degree 1, the nodes of degree 2 must be 25, or let's say 24, 25 or 26, depending on the integer rounding and on the fluctuations due to the term $\xi$. The degree sequence \texttt{list\_of\_nodes} will start with 100 elements equal to 1, then 25 elements equal to 2, and so on. This degree sequence is fed to the wiring function \texttt{configuration\_model}, which creates the ``stubs'' of the network and adds links at random between them, until each node reaches the degree defined by the degree sequence.

\section{Results of the rewiring procedure for scale-free networks}
\label{results-sf}

If we use as target the Vazquez-Weigt correlations, the average $r$ coefficient of the resulting network ensemble is very close to zero, independently from the $r$ parameter of the target. The ensemble average of $\bar{k}_{nn}$ is linearly increasing only for small values of $k$, similarly to what happens for the rewiring of BA networks. We can thus conclude that the Vazquez-Weigt assortative correlations are incompatible with a scale-free degree distribution, not only due to structural disassortativity at large degrees, but for all degrees except in a small range close to $k_{min}$.

One might wonder whether it is possible at all to obtain positive ensemble values of $r$ by rewiring a scale-free network. The answer is affirmative, as shown by Xulvi and Brunet \cite{xulvi2004reshuffling} with their ``empirical'' rewiring method (links pairs are exchanged if the differences between their degrees increase) and also in \cite{bertotti2019configuration,bertotti2020network} with a maximally assortative rewiring based on a formula for the variation $\Delta r$.

It is also possible to obtain assortative scale-free networks by assortative rewiring with correlations different from the Vazquez-Weigt recipe. This has the advantage, compared to empirical rewiring or maximally assortative rewiring, of maintaining some control about the functional form of the correlations. We shall illustrate two methods, denoted for brevity as BM1 and BM2.

\subsection{BM1 assortative matrices}

This set of matrices is built through a procedure \cite{bertotti2016bass} in which the matrix elements of $P(h|k)$ are chosen to be largest on the main diagonal and decreasing elsewhere, then properly adjusted in order to satisfy the Network Closure Condition and normalized column-by-column. By performing the Newman rewiring with this kind of matrices as target, one obtains ensembles with $r\simeq 0.2 - 0.3$ and average $\bar{k}_{nn}$ as shown in Figs.\ \ref{bm-cloud}, \ref{bm-mean}.

\subsection{BM2 assortative matrices}
\label{bm2-matrices}

An alternative procedure for constructing assortative matrices \cite{bertotti2019evaluation} is to define them with an explicit formula in terms of the excess degrees, like e.g.\ 
\begin{equation}
    e_{jk}=\frac{c_\delta}{(1+j+k)^\delta}, \qquad j,k=0,...,n-1
\end{equation}
where $c_\delta$ is a normalization constant. From this one can obtain the degree distribution $P(k)$ in the form of a series and check that it behaves with good approximation as a scale-free distribution. The $P(h|k)$ matrix is obtained from $e_{jk}$ and $P(k)$ as usual. The target $\bar{k}_{nn}$ function is, unlike in the previous cases, increasing but definitely non-linear (see Fig.\ \ref{ejk-mean}). The disadvantage of this method is that the scale-free exponent $\gamma$ cannot be fixed a priory but depends on the $\delta$ exponent in the definition of $e_{jk}$.

\section{Numerical solutions of the Bass diffusion equation on assortative rewired networks}
\label{num}

As mentioned in the Introduction, when we have a concrete realization (actually, a statistical ensemble of realizations) of a network having assigned degree distribution and correlations as close as possible to a certain theoretical target, we can solve systems of differential equations which describe a certain dynamical process unfolding on the network. The equations in the system are as many as the nodes. This means that the behavior of single agents is described much more accurately than in the HMF approximation, at the cost of course of a greater computational complexity which limits the network size to a few thousands of nodes.

The results of the numerical solutions still need to be aggregated in order to examine them. Nevertheless, one observes some interesting differences in comparison to the HMF results. To some extent these differences are due, in the scale-free case, to the real random occurrence of large hubs in the network, while in the HMF approximation all hubs contribute ``virtually'' to dynamics, each one with a small probability.

We have focused our attention on the network Bass model, which written on the single nodes in first closure \cite{newman2010networks,gleeson2011high} takes the form
\begin{equation}
    \frac{dX_l(t)}{dt}=(1-X_l(t)) \left[ p+q\sum_{j=1}^N A_{lj}X_j(t) \right], \ \ \ l=1,\ldots,N
\end{equation}
where $A$ is the adjacency matrix of the network and the variable $X_l$ has to be understood as the
expectation $\langle \xi_l \rangle$ of the non-adoption ($\xi_l=0$) or adoption ($\xi_l=1$) state of node $l$ over many stochastic evolutions of the system. The code for the solution is available at \texttt{https://github.com/Ladilu/python-bass-accessible} and makes use of the Python method \texttt{odeint} which can be imported from the package \texttt{scipy.integrate}. The adjacency matrix is handled by \texttt{networkX} as a sparse matrix, with the commands \texttt{m = nx.adjacency\_matrix(G)}, \texttt{A=m.todense()}. Like in the HMF equations, the imitation coefficient $q$ is normalized to the average connectivity $\langle k \rangle$, in such a way to allow comparisons between networks of different kind while re-scaling the dominant effect of the average degree on diffusion times.

The Bass model is reduced to a standard SI epidemic model when the publicity coefficient $p$ is set to zero; however, the $p$-term allows to produce a meaningful diffusion dynamics also starting from null initial conditions, i.e., without initial adopting nodes (the equivalent of infected nodes in the SI model). One can thus define a characteristic peak adoption time $t_{max}$ as the time at which the adoption rate $f(t)$ is maximum. $f(t)$ is the derivative of $F(t)$, which represents the fraction of total adopters as a function of time and has a typical S-shaped plot ($F(t)=\sum_{l=1}^N X_l(t)$). It is also possible to define other characteristic times, like the takeoff time, see \cite{bertotti2019evaluation}.

In general one observes that for scale-free networks which have the same critical exponent $\gamma$ and have been rewired using the same target correlations, $t_{max}$ depends strongly on the degree $k_{max}$ of the largest hub present in the network (at least for the networks with 1000 nodes examined). The presence of a super-hub makes diffusion visibly faster, as could be intuitively expected, and it also appears to make the assortative rewiring process less efficient, leading to smaller values of the final $r$ in correspondence of the same target correlations.

If we want to assess the effect of assortative correlations we thus need to compare network realizations which have similar $k_{max}$. Then it turns out that in assortative networks $t_{max}$ is systematically smaller than in uncorrelated networks; in addition, the diffusion rate reaches its peak earlier, and decreases faster after the peak. See Tabs.\ \ref{tab-figs}, \ref{tab1}, \ref{tab2}. This outcome differs from the predictions of HMF theory, according to which uncorrelated networks give in general a smaller $t_{max}$ than assortative networks \cite{bertotti2019bass}.

\begin{table}
\begin{center}
\begin{tabular}{cc}
  \includegraphics[height=0.25\textheight]{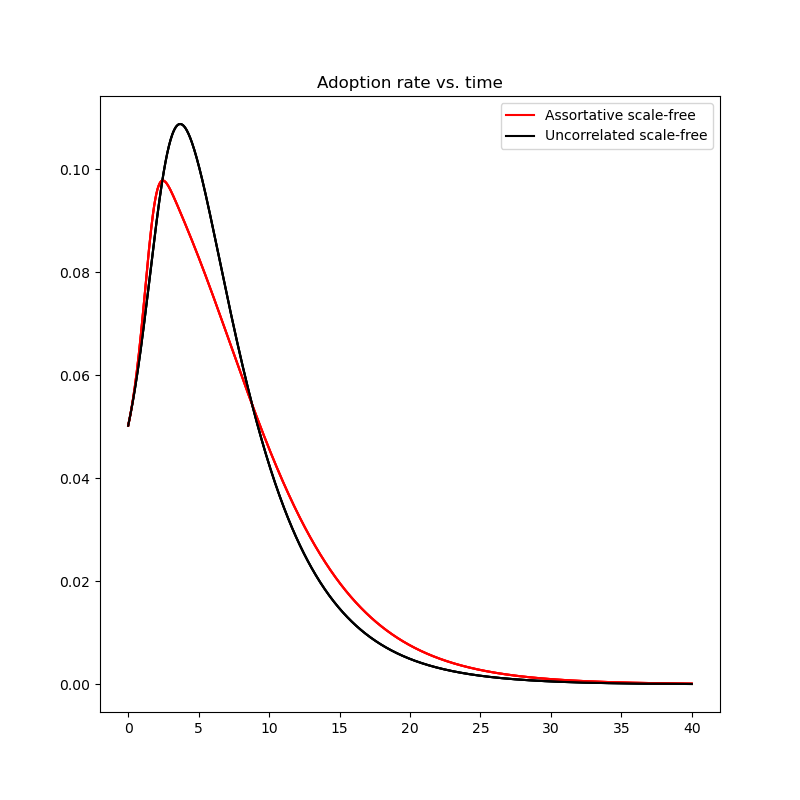}
  &
  \includegraphics[height=0.25\textheight]{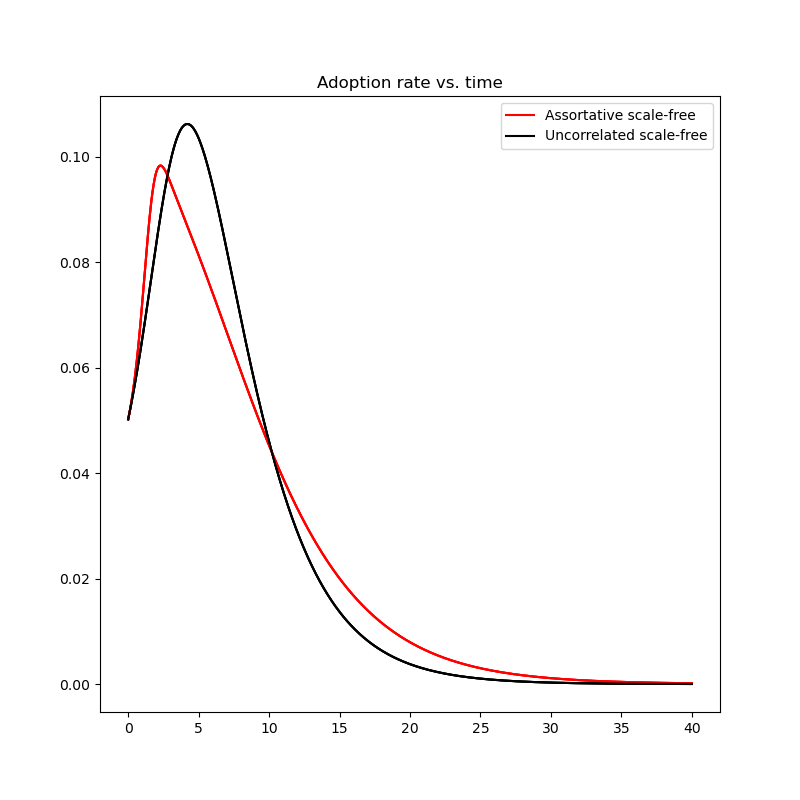}
   \\                                                     
   $\gamma = 2.5$& $\gamma=3$
\end{tabular}
\end{center}
\caption{\label{table_image}Total adoption rates as a function of time for a network Bass model running on uncorrelated and assortative scale-free networks with 1000 nodes (system of coupled differential equations for the single nodes).}
\label{tab-figs}
\end{table}

\begin{table}
\begin{center} 
\textbf{Assortative} \\
\begin{tabular}{|c|c|c|c|}
\hline
$\mathbf{k_{max}}$ & $\mathbf{<k>}$  & $\mathbf{t_{max}}$ & $\mathbf{r}$  
\\ \hline
90 & 4.2& 2.4 & 0.23  
\\ \hline
98 & 4.2 & 2.4 & 0.19
\\ \hline
132 & 4.8 & 2.0 & 0.09
\\ \hline
149 & 4.5 & 2.0 & 0.11
\\ \hline
\end{tabular}
\end{center}
\caption{Examples of peak times $t_{max}$ observed in assortative networks with correlations of the BM1 type, $\gamma=2.5$, $N=1000$, $k_{min}=2$. Note the strong dependence on the degree $k_{max}$ of the largest hub, as discussed in the text.} 
\label{tab1}
\end{table}

\begin{table}
\begin{center} 
\textbf{Uncorrelated} \\
\begin{tabular}{|c|c|c|c|}
 \hline
$\mathbf{k_{max}}$ & $\mathbf{<k>}$  & $\mathbf{t_{max}}$ & $\mathbf{r}$  
\\ \hline
30 & 3.1& 4.2 & -0.03  
\\ \hline
33 & 3.1 & 4.1 & 0.02
\\ \hline
122 & 4.3 & 3.2 & -0.08
\\ \hline
126 & 4.3 & 3.1 & -0.08
\\ \hline
\end{tabular}

\end{center}
\caption{Same as in Tab.\ \ref{tab1}, but with uncorrelated networks. For comparable $k_{max}$, the peak times are larger than with assortative networks. In the rewiring code, a Vazquez-Weigt matrix with $r=0$ is chosen as target. The resulting effective $r$ is slightly negative due to structural disassortativity.
} 
\label{tab2}
\end{table}

\section{Outlook: NetLogo simulations}
\label{Netlogo-sim}

\begin{figure}[t!]
\centering
\begin{subfigure}[t]{\linewidth}
        \includegraphics[width=40mm]{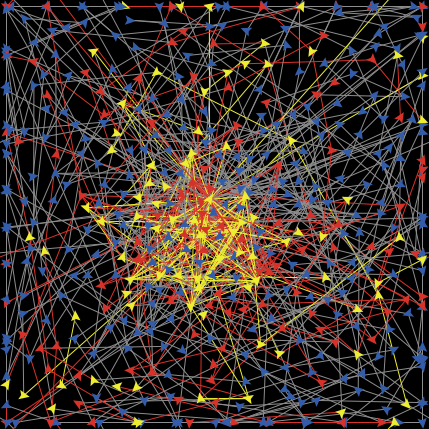}
        \label{fig:a}
\end{subfigure}
~
\begin{subfigure}[t]{\linewidth}
        \includegraphics[width=40mm]{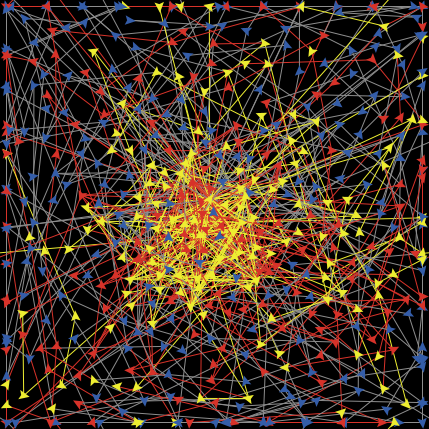}
        \label{fig:b}
\end{subfigure}
\begin{subfigure}[t]{\linewidth}
        \includegraphics[width=40mm]{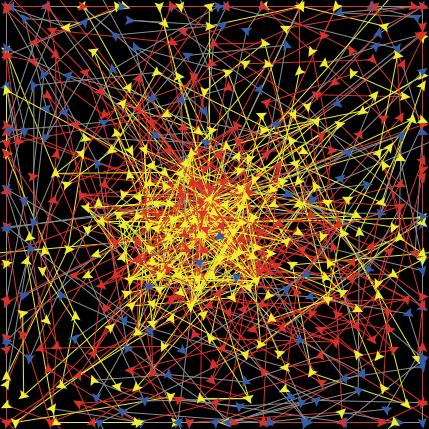}
        \label{fig:c}
\end{subfigure}
\begin{subfigure}[t]{\linewidth}
        \includegraphics[width=40mm]{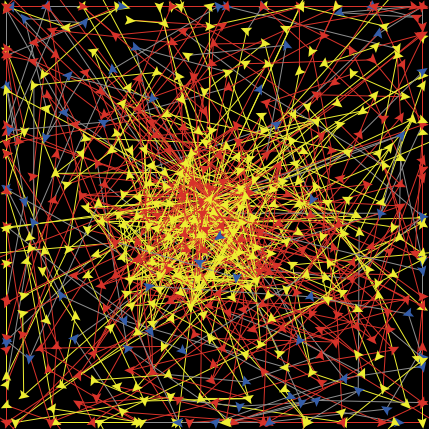}
        \label{fig:d}
\end{subfigure}
\begin{subfigure}[t]{\linewidth}
        \includegraphics[width=40mm]{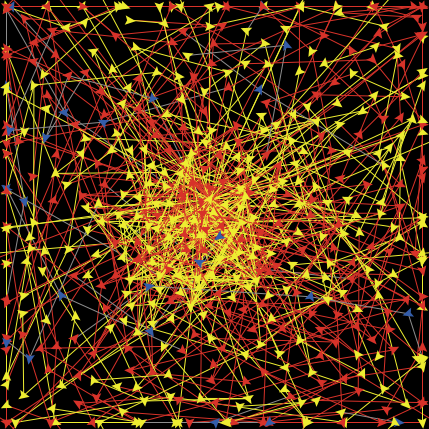}
        \label{fig:e}
\end{subfigure}
\begin{subfigure}[t]{\linewidth}
        \includegraphics[width=40mm]{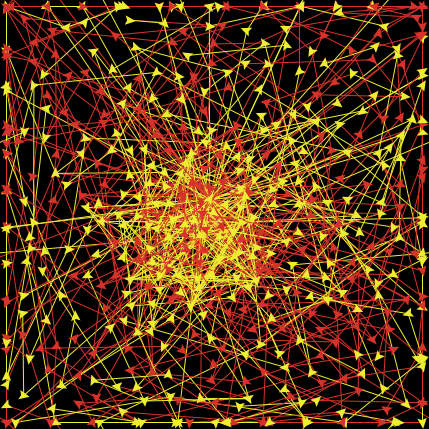}
        \label{fig:f}
\end{subfigure}
\caption{Some \texttt{NetLogo} views of the agent dynamics for Bass adoption in an assortative scale-free network. The agents that have not adopted are initially colored in blue, then ``broadcast'' (publicity) adoptions are represented by red color and ``network'' (imitation) adoptions by yellow color. The network has 500 nodes, $\gamma=2.5$ and is obtained via Newman rewiring with a BM1 target matrix. }
\label{fig:NetLogo}
\end{figure}

In order to reconstruct dinamically the evolution of the Bass diffusion on an assortative network, preliminary agent-based simulations have been carried out with \texttt{NetLogo}. The code is also available at \texttt{https://github.com/Ladilu/python-bass-accessible} and it consists of a few procedures. \texttt{NetLogo} allows the uploading of networks created by means of applying  the Python rewiring algorithm with the criterium  of setting a target BM1  matrix to an initial graph $G$. The graph $G$ is built costructing the degree sequence from a power law and then applying the \texttt{configuration\_model} procedure. The Python code is able to produce a \texttt{graphml} output for the nodes and links of the network . There exists a corresponding method in the \texttt{NetLogo} code , \texttt{wire\_graphml}, which loads the graphml file and then the procedure \texttt{create-network} uses the method to effectively reconstruct the corresponding network. Another method in the procedure \texttt{create-network} is \texttt{setup-nodes}, which places the network in the screen within the area of a circle so that the coordinates of the points corresponding to the agents are well readable. There exists the possibility of setting an arbitrary number of seed adopters. In the procedure \texttt{adopt} the Bass model is implemented. The $p$ parameter for the broadcast influence is compared to a random variable, so that according to its value the agents which are subject to publicity adoptions are given the red color.  The $q$ parameter  for a given node is set using the social influence parameter normalized to the number of neighbors divided by the mean degree of the network, which is computed separately. The \texttt{adopt} procedure is then used in the \texttt{go} procedure that can be set in the interface to make the diffusion process start in the simulation. The time-based approach is chosen, by means of resetting the ticks count before any new start of the Bass model simulation. The process can be monitored on the screen as represented in Fig. \ref{fig:NetLogo}. A detailed description of the simulations and their results will be presented in a forthcoming paper.

\bibliographystyle{ieeetr}
\bibliography{nets}

\end{document}